\documentstyle[12pt,epsfig]{plenum}

\title{UNIVERSAL PROPERTIES OF ANGULAR \\CORRELATIONS IN
QCD JETS}
\author{Wolfgang Ochs\\ \\
Max-Planck-Institut f\"ur Physik\\
F\"ohringer Ring 6, D-80805 M\"unchen, Germany
}
\def\t12{\theta_{12}}
\def\theta{\vartheta}

\def\vth{\vartheta}

\def\phi{\varphi}
\def\col{c_a}
\def\ve{\varepsilon}
\def\epsilon{\varepsilon}

\def\rotwo{\rho^{(2)}}

\def\as{\alpha_s}

\def\epem{e^+e^-}
\begin{document}

%Here is the title page for the preprint
\mbox{ }\\
\hspace*{110mm} MPI-Ph/94-68\\
\hspace*{110mm} September 1994\\
\vfill
\begin{center}
{\bf UNIVERSAL PROPERTIES OF ANGULAR CORRELATIONS IN} \\
{\bf QCD JETS}
\footnote{To be published in the Proceedings of the Workshop
 ``Hot Hadronic matter: Theory and experiment'', Divonne-les-Bains,
  France, Plenum, N.Y.}\\
\vfill
WOLFGANG OCHS\\
\vspace{0.5cm}
{\sl
Max Planck Institut f\"ur Physik \\
Werner-Heisenberg-Institut\\
F\"ohringer Ring 6, D-80805 M\"unchen} \\
\vfill%\vspace{2.5cm}
{\bf Abstract}\\
\vspace{5mm}
\end{center}
\noindent Predictions
for angular correlations between an arbitrary number of partons
are derived in the high energy limit. The quantities considered depend
on angles and primary energy through a single variable $\varepsilon$
which
implies certain scaling properties and relations
between quite different observables. These asymptotic predictions
derived in the double log approximation of QCD
are checked against Monte Carlo calculations at
the parton and hadron level.
\vfill%
\mbox{ }
% Here starts the regular paper for proceedings
\newpage
\maketitle
{\protect\abstract Predictions
for angular correlations between an arbitrary number of partons
are derived in the high energy limit. The quantities considered depend
on angles and primary energy through a single variable $\ve$ which
implies certain scaling properties and relations
between quite different observables. These asymptotic predictions
derived in the double log approximation of QCD
are checked against Monte Carlo calculations at the parton and hadron level.
\endabstract}
\section*{INTRODUCTION}
\hspace*{\parindent} Many
new results from the detailed study of multiparticle
production processes  have been obtained in recent years,
still there is no fully satisfactory model to
describe all these phenomena. One approach is based on perturbative QCD,
particularly suited for hard processes
involving large momentum transfers,
in which an initially scattered parton evolves by gluon Bremsstrahlung
into a jet of partons and finally into the observable hadrons,
whereby the effect of the color confinement force is modeled.
Another approach relies on a statistical and thermodynamic description in
which the microscopic degrees of freedom are integrated over and only
global quantities are kept. Hadrons are produced from a quark gluon
plasma after a phase transition. This approach is applied in particular
to nuclear collisions in the search for the quark gluon plasma.
As many aspects of particle production change only moderately when going
from the more elementary collisions like
$\epem$ annihilations to the more complex $pp$ or nuclear collisions
it is desirable to develop both the perturbative and
the statistical methods and to explore their predictive power.

In any case, for a satisfactory model of multiparticle production we
would like to have finally a mathematical model based on a simple
principle, there should be only a few parameters, some results should
be obtained analytically -- even if only approximately --
otherwise the structural properties of the theory can hardly be fully
explored. A nice example of a model of this type is Hagedorn's bootstrap model:
it is based on a single guiding principle, the bootstrap
principle, the only parameters being the particle masses and the
interaction volume; analytical results on multiparticle
observables can be derived from an equation for the
generating functional of multiparticle densities\cite{HM}.
Although it has been attempted to extend
the statistical approach to hard processes \cite{OC}
the most convincing results for such processes
are obtained today from the parton cascade model
as derived from perturbative QCD.

In this talk I would like to report on results from
perturbative QCD on angular correlations inside a parton jet
which are obtained in collaboration with Jacek Wosiek
\cite{OW1,OW2}.
The essential parameters of the parton cascade are the
scale parameter $\Lambda$,
which determines the coupling strength and a
cutoff $Q_o$ which regulates the infrared and collinear singularities
of the gluon Bremsstrahlung. As to the hadronization
process we follow here the idea that it is sufficiently
soft and the distribution of hadrons follows largely the
distribution of partons \cite{SOFT}. In particular for momentum
spectra of partons and hadrons such an equality up to a constant
has been established, if the parton cascade is evolved down to the
hadronic (i.e.\ pion) mass scale $Q_o\approx m_h$ \cite{LPHD}.
Such a similarity may also be expected if the observable
considered is ``infrared save'', i.e.\ does not depend
explicitly on the cutoff $Q_o$. It is one of the
motivations of the present study, as to what
extent such a hypothesis of soft hadronization is
actually correct and supported by the experiment.
It would allow for a certain class of observables to
be calculated just only in terms of $Q_o$ and $\Lambda$ parameters,
resulting in a scheme of high predictive power.

\section*{THEORETICAL SCHEME}
\hspace*{\parindent} Our
calculations are based on the double logarithmic
approximation (DLA) of QCD. The probability to
emit a single gluon of momentum $K$ from a primary parton
$a$ of momentum $P$ is given by
\begin{equation}
 {\cal M}_{P,a}(K)d^3 K=\col a^2(K\Theta_{PK})\frac{dK}{K}
\frac{d\Theta_{PK}}{\Theta_{PK}}\frac{d\Phi_{PK}}{2\pi},
\label{adef}
\end{equation}
where $a^2\equiv\gamma^2_0=6\alpha_s/\pi$ is the QCD anomalous
dimension of the multiplicity evolution and $c_g=1$ or $c_q=4/9$
for gluon or quark jets respectively. For the running coupling
constant we write $a^2(p_T)=\beta^2/(\ln (p_T/Q_0)+\lambda)$
whereby $\lambda=\ln (Q_0/\Lambda)$ and $\beta^2=12(\frac{11}{3}
N_c -\frac{2}{3} N_f)^{-1}\approx 1.565$ for 5 flavours $N_f$.
In this approximation the recoil effects are neglected (i.e.\
energy and momentum conservation are violated), integrals are performed by
retaining only the leading contributions from collinear and
soft divergencies. The interference of the soft gluons is taken
into account by the ``angular ordering'' prescription. In DLA
one obtains the leading asymptotic behaviour, but the non-asyptotic
corrections of relative order $\sqrt{\as}$ are potentially large.

The generating functional which includes the
``leading'' primary parton in the final state is given by the
non-linear integral equation \cite{FAD}
\begin{equation}
   Z_{P,a}\{u\}= u(P)
\exp \left( \int_{\Gamma_P(K)} {\cal M}_{P,a}(K) [u(K) Z_{K,g}\{u\} -1]
d^3 K \right).
\label{mez}
\end{equation}
In this form the virtual corrections are included which ensures the
proper normalization $Z_P\{u\}\Big|_{u=1}=1$. The density
distribution of $n$ partons is then obtained by
functional differentiation after the test functions
$u(k)$
\begin{equation}
\rho^{(n)} (k_1,...,k_n)= \delta^n Z\{u\}/\delta u(k_1)...
\delta u(k_n)\mid_{u=1} .     \label{den}
\end{equation}
Similarly, the cumulant (connected) correlation function
is derived as in Eq.(\ref{den}) but with $Z$ replaced by
$\ln Z$. Starting with $n=1$ one obtains from (\ref{den})
and (\ref{mez}) linear integral equations which can be solved
recursively. Here we specialize on angular distributions
and the corresponding equations are obtained after integration
over the momenta. We have obtained results \cite{OW1,OW2} on the general
inclusive cumulant correlation function of $n$ particles
in their spherical angles. Of special interest are the distribution
in the relative polar angle $\t12$ between two partons
and the multiplicity moments of general order $n$
for particles falling into sidewise angular regions.

It turns out that all these angular observables can be
constructed from a generic function $h^{(n)}(\delta,\vartheta, P)$
which fulfils the integral
equation
\begin{equation}
h^{(n)}(\delta,\theta,P)=
d^{(n)}(\delta,\theta,P)+\int_{Q_0/\delta}^{P} \frac{dK}{K}
\int_{\delta}^{\theta}
\frac{d \Psi}{\Psi} a^2 (p_T) h^{(n)}(\delta,\Psi,K).  \label{hint}
\end{equation}
where $\delta$ and $\vartheta$ are the small and the large
angles of the respective problem and the inhomogeneous term behaves at high
energies like
%\begin{equation}
$ d^{(n)}\sim \exp (2n\beta\sqrt{\ln(P\delta/\Lambda)})$.
%\label{inhom}
%\end{equation}
One finds that the natural variables of the problem, instead of
$P,\vartheta$ are
\begin{equation}
\epsilon =\ln (\theta/\delta)/\ln (P\theta/\Lambda)\/, \qquad
\zeta=1/(\beta\sqrt{\ln (P\theta/\Lambda)}).
\label{epszet}
\end{equation}
As $\vth>\delta\geq Q_0/P>\Lambda/P$, we find $0\leq\ve\leq 1$.
The solution for $\ln h^{(n)}$ can be written as an expansion in
$\zeta\sim\sqrt{\alpha_s}$.
In the high energy limit ($\epsilon$ fixed, $P\to\infty)$
one obtains
\begin{equation}
h^{(n)}(\delta,\vartheta,P)\sim {{\rm exp}}
(2\beta\sqrt{\ln (P\vartheta/
 \Lambda)}\omega (\epsilon, n)).
\label{hasy}
\end{equation}
The scaling function $\omega(\epsilon,n)$ is known in implicit form
\cite{OW1}. For small $\epsilon$ it has a power expansion
%\begin{equation}
%\omega(\epsilon,n) = n - \frac{1}{2} \frac{n^2-1}{n}\epsilon + \ldots
%~\bigl(1 + \frac{n^2+1}{4n^2} \epsilon + \frac{1}{8n^4}
%(n^4 + \frac{4}{3} n^2 + \frac{5}{3}) \epsilon^2 + \cdots\bigr)
%\label{omex}
%\end{equation}
$\omega(\epsilon,n) = n - (n^2-1)\epsilon/2n + \ldots$
Another useful approximation obtains from an expansion in $1/n$ which yields
%\begin{equation}
%\omega(\epsilon,n)\approx n \sqrt{1-\epsilon}~(1 - 1/(2n^2) \ln
%(1-\epsilon))
%\label{omln}
%\end{equation}
$\omega(\epsilon,n)\approx n \sqrt{1-\epsilon}~(1 - \frac{1}{2n^2} \ln
(1-\epsilon))$
and has a 1\% accuracy for $\epsilon < 0.95$ already for $n=2$.
An asymptotic behaviour of type (\ref{hasy}) was also found in the study of
azimuthal particle correlations \cite{DMO}.

\section*{POLAR ANGLE CORRELATIONS}
\hspace*{\parindent} First
we discuss the correlations  $ \rotwo(\t12,P,\Theta)\equiv dn/d\t12$
in the relative polar angle
$\vartheta_{12}$ of two partons both inside a forward cone
around the initial parton of half opening angle $\Theta$.
For the study of
scaling properties it is more convenient to consider the
distribution in the variable $\varepsilon=\ln(\Theta/\vartheta_{12})/
\ln(P\Theta/\Lambda)$ given by $\hat\rho^{(2)}(\epsilon)\equiv
dn/d\epsilon = \vartheta_{12}\ln(P\Theta/\Lambda)\rotwo(\t12)$. For
the correlation $\hat r(\epsilon)=\hat\rho^{(2)}(\epsilon)/
\bar n^2$ normalized by the multiplicity in the
cone\footnote{this normalization has
better scaling properties \cite{OW2}
than the differential one used earlier\cite{OW1}}
$\bar n$
one obtains in the high energy limit ($\epsilon$ fixed,
$P\to\infty)$ from Eq.\ (\ref{hasy}) with
$h^{(2)}(\vartheta_{12})\sim \vartheta_{12}\rho^{(2)}(\vartheta_{12})$
for either quark or gluon jet
 \begin{equation}
 \hat r(\epsilon)=2\beta\sqrt{\ln(P\Theta/\Lambda)} \exp \left(-2\beta
 \sqrt{\ln(P\Theta/\Lambda)} \left(2-\omega(\epsilon,2)\right)
    \right).
\label{rhat}
\end{equation}
Differences between quark and gluon jets and the influence
of the leading particle show up at finite energies where we
obtain
\begin{eqnarray}
\hat{r}_a(\epsilon)&=&c_a^{-1}y\exp
(-y(2-\omega(\epsilon,2))
   -   (c_a^{-1}-1)y\exp(-2y(1-\sqrt{1-\epsilon})) \nonumber  \\
 & + & 2\beta\sqrt{2y}/(c_af(1-\epsilon)^{{1\over 4}})
        \exp (-y(2-\sqrt{1-\epsilon}))
\label{rhqg}
\end{eqnarray}
with $y=2\beta\sqrt{\ln(P\Theta/\Lambda)}$,
$f=2\beta K_0(2\beta\sqrt{\lambda})/\sqrt{\pi} \approx 0.145$
for $\lambda =\ln 2$. To explore the
scaling properties of $\hat r(\ve)$ we consider the quantity
$-\ln \hat r(\ve)/(2\sqrt{\ln(P\Theta/\Lambda)})$
depending only on $\ve$ for any momentum $P$ or
cone opening angle $\Theta$. The asymptotic limit and a finite
energy result are shown in Fig.~1.
% Fig.1 was first here
%
As a test of our analytical calculations we also compared to the
Monte Carlo evaluation of the parton cascade.\footnote{we
used the program HERWIG \cite{HERWIG} with parameters
$\Lambda=0.15~{\rm GeV}$, $m_q=m_g=0.32$~GeV and without
non-perturbative gluon splitting for the process $\epem\to u\bar u$.}
As can be seen from the
figure, the predicted asymptotic scaling behaviour is nicely
reproduced for small $\ve\leq 0.5$ in a large energy range
$(P\geq 20$~GeV) whereas deviations occur for larger $\ve$
(smaller relative angles $\t12$ approaching the cut-off $Q_0/P$).
It should be noted that the
normalization of the above expressions in (\ref{rhat},
\ref{rhqg}) and in particular the quark and gluon
difference at finite energy are of non-leading order in the DLA
and therefore less reliable. We have therefore adjusted
the overall normalization of the functions in (\ref{rhat},
\ref{rhqg}) to the Monte Carlo data.
We also studied the effect of hadronization provided by the
HERWIG MC. Again these effects are negligable for small
$\ve\leq 0.5$ \cite{OW2}.
Preliminary data from DELPHI \cite{DELPHI} have recently given first evidence
for approximate $\ve$-scaling in $\Theta$ $(\Theta\geq 30^o)$ and a close
proximity of the data to the parton model results.

\section*{MULTIPLICITY MOMENTS FOR SIDEWISE ANGULAR CONE OR RING}
\hspace*{\parindent} In
a second application we consider particle multiplicities in
a sidewise cone $\delta\Omega$ of half opening $\delta$ at
polar angle $\vartheta$ with respect to $\vec P$
and  in an angular
ring of width $2\delta$ symmetrically around the primary
parton direction, centered at polar angle $\vartheta$.
We will refer to the ring and the cone as the 1D and 2D configurations.
We calculate the factorial and cumulant multiplicity
moments $f^{(n)}$ and $c^{(n)}$
or normalized by the  multiplicities $\bar n$ in the respective angular regions
$F^{(n)}=f^{(n)}/\bar n^n$ and
$C^{(n)}=c^{(n)}/\bar n^n$ ($f^{(2)}=<n(n-1)>$, $C^{(2)}=F^{(2)}-1$, etc.).
The cumulant moments $c^{(n)}$ can
be derived again from the generic equation
(\ref{hint}).
%with (\ref{inhom}).
For the normalized moments
we find
\begin{equation}
C^{(n)}(\vth,\delta) \sim (\vth/\delta)^{D(n-1)}
\exp\left(-2\beta\sqrt{\ln(P\theta/\Lambda)}(n-\omega(\ve,n))\right)
\label{cmomno}
\end{equation}
with $\ve=\ln(\vth/\delta)/\ln(P\vth/\Lambda)$. The
dependence on $\ve$ is as in
$\hat r(\ve)$ discussed above.
Similar results have also been found by other groups
\cite{DM,BMP}. In the modified LLA including next to
leading effects in the exponent
\cite{DM} the power changes typically by
10\%.
\begin{figure}[t]
%\centerline{\mbox{\epsfig{figure=fig1.ps,hight=132mm}}}
\centerline{\mbox{\epsfig{figure=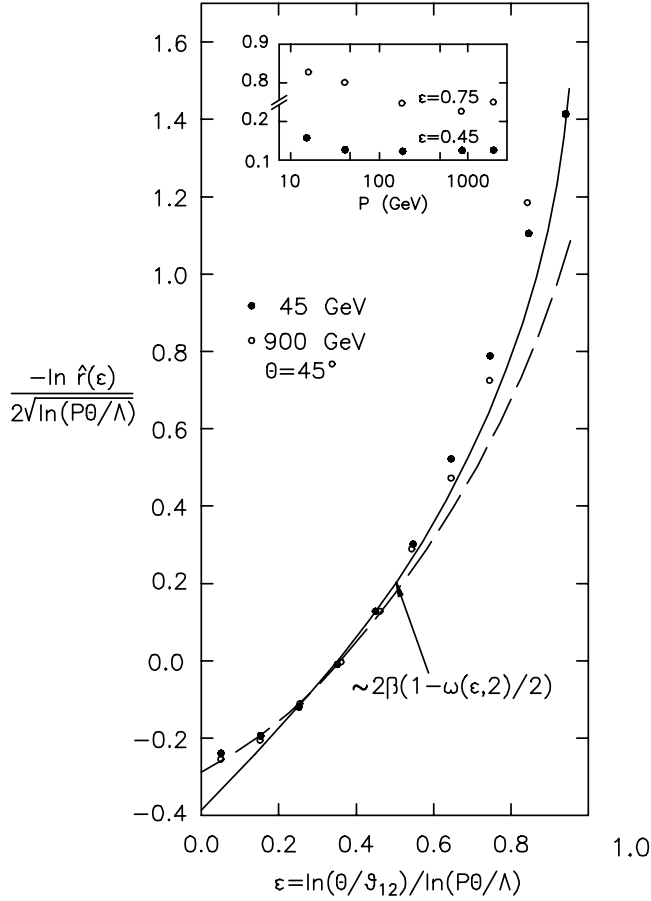,width=85mm}}}
\caption{\protect\small Rescaled
polar angle correlation function for the high
energy limit, Eq. (7) (full line), and for a quark jet
of 45~GeV, Eq. (8) (dashed line), with normalization adjusted to the data
as obtained from the HERWIG MC
at the parton level.
The insert shows the energy dependence of the same quantity for
fixed $\ve$.
}
%\protect\label{f1}
\label{fig:f1}
\end{figure}

The scaling properties of the various moments can again be
conveniently investigated by projecting out the $\ve$-dependence of the
exponent in
Eq.(\ref{cmomno}), one finds for
\begin{equation}
-\hat C^{(n)} \equiv  -{\ln \lbrack(\delta/\theta)^{D(n-1)}
C^{(n)}\rbrack \over n\sqrt{\ln (P\theta/ \Lambda)}}
 =  2\beta (1-{\omega(\epsilon ,n)\over n})
\label{chat}
\end{equation}
in the high energy limit, or $-\hat C^{(n)}
 \simeq  2 \beta(1-\sqrt{1-\epsilon})$
for large $n$ independent of n,D.
%(see Eq. (\ref{omln})).
\begin{figure}[t]
%\vspace{10.2cm}
%\centerline{\mbox{\epsfig{figure=fig2.ps,hight=102mm}}}
\centerline{\mbox{\epsfig{figure=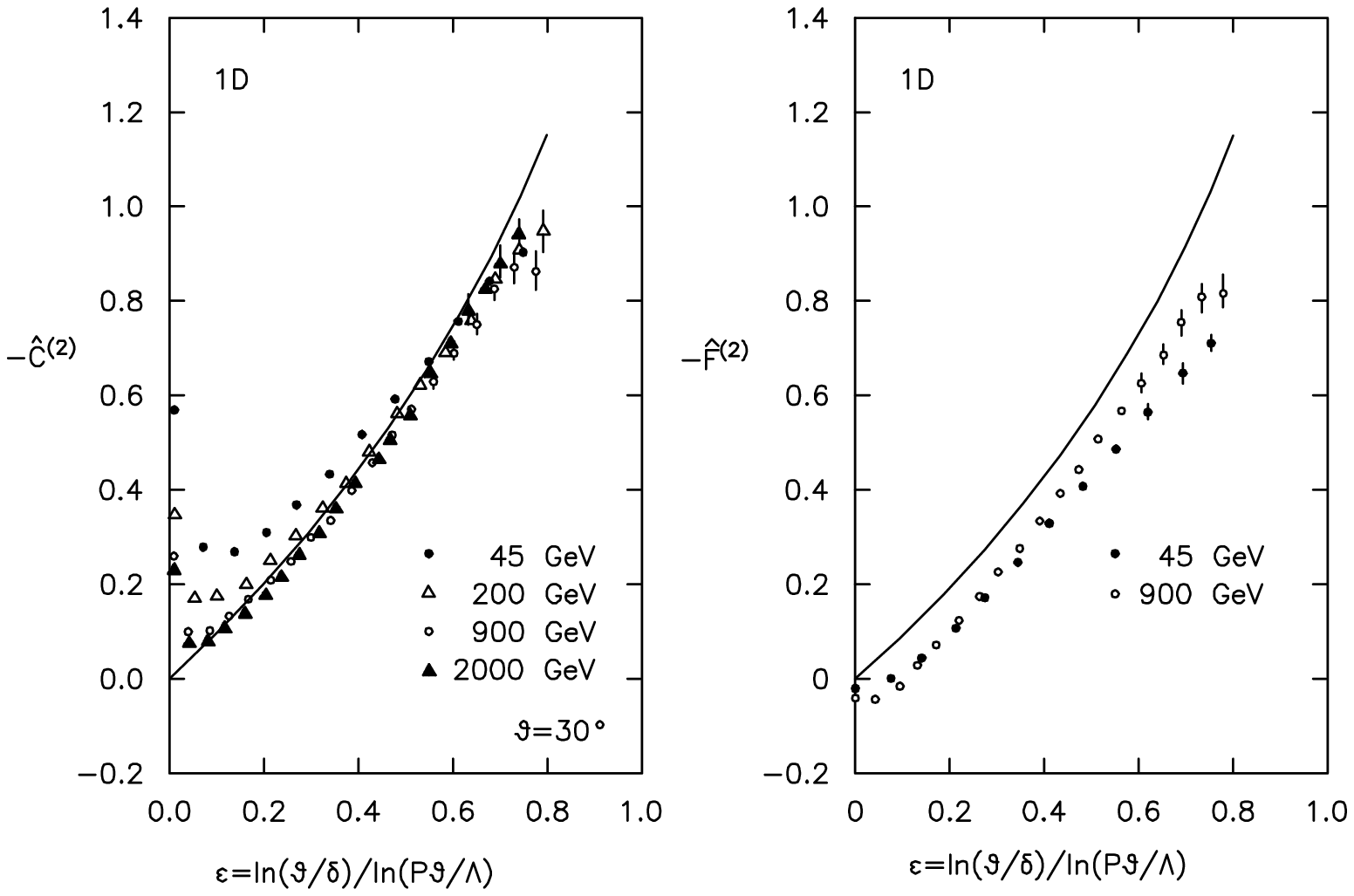,width=160mm}}}
\caption{\protect\small
(a) Rescaled cumulant moments for the ring as defined in
Eq.~(10) for the parton MC for different jet momenta $P$
in comparison with the asymptotic prediction Eq. (10); (b)~same
as (a) but for factorial moments.}
\label{fig:f2}
\end{figure}

In Fig.~2a we plot $-\hat C^{(2)}$ for the ring $(D=1)$ vs.\
$\ve$ for different primary momenta $P$ for the parton MC.
There is a violation of $\ve$-scaling for small $\ve$ but the
asymptotic prediction is approached for high energies. We have
repeated the same calculation as in
 (\ref{chat}), but for the factorial moments $F^{(n)}$ replacing $C^{(n)}$
in (\ref{chat}), see Fig.~2b.
As $F^{(2)}=C^{(2)}+1$ they approach the same asymptotic limit.
Apparently, the nonasymptotic corrections are such that the scaling
in $\ve$ for small $\ve$ sets in already at low energies for the
factorial moments. The results for higher moments follow roughly
the expectation (\ref{chat}),
on the other hand the $D=2$ moments show a
more gradual dependence on $\ve$ than predicted \cite{OW2}.

An interesting aspect of these results is the remarkable
universality of the various moments (different $n,D$) and also of
the very different observable $\hat r(\ve)$ which all converge
against the same limiting function after appropriate rescaling (see Figs. 1,2).

\section*{SUMMARY}
\hspace*{\parindent} The
angular observables considered here, after appropriate
rescaling, approach a limit in the normalized logarithmic
angular variable $\ve =\ln(\vth/\delta)/\ln (P\vth/\Lambda)$
for $\ve$ fixed, $P\to\infty$. The comparison with the
parton MC shows that this limit is already approached for
2-particle correlation $\hat r(\ve)$ and the factorial
moments $F^{(n)}$ at present energies sufficiently far away
from the angular cutoff $Q_0/P$
$(\ve\leq 0.5)$ whereas the cumulant
moments approach the asymptotic limit only at higher energies
$(P\sim 1$~TeV). In the region of small $\ve$ -- not further discussed
here -- the observables approach a power behaviour in the angular variables
and become independent of the cutoff $Q_0$ ("infrared safe").
It is in this region
that also the hadronisation corrections are found small.

The experimental confirmation of the universal
high energy behaviour of rather different
angular observables and
of $\ve$-scaling with two redundant variables $P$ and $\vth$
can provide further evidence for a soft confinement mechanism and
parton hadron duality which allows
to calculate hadronic distributions directly from the partonic
ones.

\subsection*{Acknowledgement}
\hspace*{\parindent} I would
like to thank R. Hagedorn for the discussions on ref. 2
and J. Wosiek for the collaboration on the
topics presented here and many discussions.

\end{document}